\documentclass[12pt]{article}
\usepackage{amsmath,amsfonts,amssymb}
\usepackage[pdftex,colorlinks=false]{hyperref}

\makeatletter
\def\numberbysection{\@addtoreset{equation}{section}
    \def\theequation{\thesection.\arabic{equation}}}
\makeatother

\numberbysection

\newcommand{\be}{\begin{eqnarray}}
\newcommand{\ee}{\end{eqnarray}}
\newcommand{\non}{\nonumber}
\newcommand{\id}{\mathbb{I}}
\newcommand{\tr}{\mathop{\rm tr}\nolimits}

\newcommand{\g}{\mathop{\mathfrak{g}}\nolimits}
\newcommand{\h}{\mathop{\mathfrak{h}}\nolimits}

\begin{document}

\begin{titlepage}
\strut\hfill UMTG--263
\vspace{.5in}
\begin{center}

\LARGE Bethe ansatz equations for open spin chains\\
\LARGE from giant gravitons\\
\vspace{1in}
\large Rafael I. Nepomechie \footnote{nepomechie@physics.miami.edu}\\[0.8in]
\large Physics Department, P.O. Box 248046, University of Miami\\[0.2in]  
\large Coral Gables, FL 33124 USA\\

\end{center}

\vspace{.5in}

\begin{abstract}
   We investigate the open spin chain describing the scalar sector of
   the $Y=0$ giant graviton brane at weak coupling.  We provide a
   direct proof of integrability in the $SU(2)$ and $SU(3)$ sectors by
   constructing the transfer matrices.  We determine the eigenvalues
   of these transfer matrices in terms of roots of the corresponding
   Bethe ansatz equations (BAEs).  Based on these results, we propose
   BAEs for the full $SO(6)$ sector.  We find that, in the
   weak-coupling limit, the recently-proposed all-loop BAEs
   essentially agree with those proposed in the present work.
\end{abstract}

\end{titlepage}

\setcounter{footnote}{0}

\section{Introduction}\label{sec:intro}

Integrability of planar ${\cal N}=4$ Yang-Mills has been investigated
primarily for closed spin chains associated with long single-trace
operators.  Indeed, the anomalous dimensions of these operators are
given by a set of closed-chain Bethe ansatz equations (BAEs)
\cite{MZ}-\cite{BS2}, which can be derived from the all-loop bulk
$S$-matrix \cite{St}-\cite{dL}. (These results are valid only 
asymptotically. For operators of finite length, there 
are finite-size corrections. See e.g. \cite{finitesize} and 
references therein.)

Nevertheless, progress is also being made in understanding open spin
chains associated with certain determinant-like operators \cite{dets}
corresponding to open strings attached to maximal giant gravitons
\cite{giants}.  This problem was first considered by Berenstein and
V\'azquez \cite{BV}, who determined the open-chain Hamiltonian
describing the mixing of such operators at one loop.  They also
computed the one-loop boundary $S$-matrix, and argued that the
Hamiltonian is integrable.  This work was extended to two loops by
Agarwal \cite{Ag} (see also \cite{OY}) and by Hofman and Maldacena
\cite{HM}.  The latter also proposed all-loop boundary $S$-matrices,
up to scalar factors which were subsequently found in \cite{CC, ABR}.
These boundary $S$-matrices were further investigated in
\cite{AN}-\cite{CY}.  \footnote{For further applications of integrable
open spin chains in gauge theory and string theory, see e.g.
\cite{DKM}-\cite{DK} and references therein.}

Although Berenstein and V\'azquez \cite{BV} found an integrable
Hamiltonian, they did not give the corresponding open-chain BAEs.
Remarkably, this problem (which is the open-chain version of the
problem solved in the pioneering work of Minahan and Zarembo
\cite{MZ}) has remained unsolved for more than four years.  One of the
aims of this paper is to determine these BAEs, whose solutions give
the anomalous dimensions of the determinant-like operators.  (As in
the periodic case, these results are expected to be valid only
asymptotically.)  Another aim of this work is to construct and
diagonalize the corresponding commuting transfer matrix (i.e., the
generating functional for the Hamiltonian and higher local conserved
charges), which would provide a direct proof of the model's
integrability.  Moreover, we would like to compare the one-loop BAEs
with the interesting all-loop BAEs which have recently been derived by
Galleas \cite{Ga} (see also \cite{MS}-\cite{MN}) from the all-loop
bulk and boundary $S$-matrices.

Berenstein and V\'azquez restricted their attention to the scalar
sector of ${\cal N}=4$ Yang-Mills, which has $SO(6)$ symmetry.  In
terms of the complex scalar fields $W, Z, Y$, Hofman and Maldacena
\cite{HM} considered operators with a large number of $Z$'s, and
distinguished two cases of interest :
\begin{description}
    \item[$Y=0$ brane:] The vacuum corresponds to the operator
    \be
    \epsilon^{j_{1}\cdots j_{N}}_{i_{1} \cdots 
    i_{N}}Y^{i_{1}}_{j_{1}} \cdots Y^{i_{N-1}}_{j_{N-1}}(Z \cdots 
    Z)^{i_{N}}_{j_{N}} \,;
    \label{Y0vacuum}
    \ee
    and excitations correspond to replacing some of the $Z$'s by 
    impurities, e.g.,
    \be
    \epsilon^{j_{1}\cdots j_{N}}_{i_{1} \cdots 
    i_{N}}Y^{i_{1}}_{j_{1}} \cdots Y^{i_{N-1}}_{j_{N-1}}(Z \cdots 
    Z \chi Z \cdots Z)^{i_{N}}_{j_{N}} \,. 
    \label{Y0excitation}
    \ee
    The sets of fields inside $( \cdots )^{i_{N}}_{j_{N}}$ constitute the
    states of the open spin chain.  The fields at the boundaries of
    this chain cannot be $Y$'s, since the operator would then
    factorize into a determinant and a single trace, and
    therefore would not describe an open string.
    
    \item[$Z=0$ brane:] The vacuum corresponds to the operator
    \be
    \epsilon^{j_{1}\cdots j_{N}}_{i_{1} \cdots 
    i_{N}}Z^{i_{1}}_{j_{1}} \cdots Z^{i_{N-1}}_{j_{N-1}}(\chi Z \cdots 
    Z \chi')^{i_{N}}_{j_{N}} \,,
     \ee
     where $\chi$ and $\chi'$ are boundary degrees of freedom; and
     excitations correspond to replacing some of the $Z$'s by
     impurities, e.g.,
     \be 
    \epsilon^{j_{1}\cdots j_{N}}_{i_{1} \cdots 
    i_{N}}Z^{i_{1}}_{j_{1}} \cdots Z^{i_{N-1}}_{j_{N-1}}(\chi Z \cdots 
    Z \chi'' Z \cdots Z \chi')^{i_{N}}_{j_{N}} \,.
    \ee

    \end{description}

For simplicity, we consider in this paper only the former case of the
$Y=0$ giant graviton brane.  Owing to the difficulty of treating
directly the full $SO(6)$ scalar sector, we instead proceed by first
examining simpler subsectors, namely, $SU(2)$ and $SU(3)$.  We then
use the results for these subsectors to conjecture BAEs for the full
$SO(6)$ sector.  Finally, we compare these one-loop BAEs with the
weak-coupling limit of the all-loop BAEs \cite{Ga}.  We find that
these two sets of results essentially agree -- the only mismatch is in
the exponent of the term corresponding to the ``massive'' node, which
also occurs in the periodic case.  \footnote{In earlier versions of 
our paper, we pointed out some errors in the first version
of \cite{Ga}, which were subsequently corrected in the second 
(published) version.}
    
The outline of this paper is as follows.  In Sec.  \ref{sec:SU3} we
consider the $SU(3)$ sector.  We construct the transfer matrix and
determine its eigenvalues in terms of roots of corresponding BAEs.
Based on these results, and on the results for the $SU(2)$ sector
which we briefly discuss in Appendix \ref{sec:SU2}, we propose in Sec.
\ref{sec:SO6} BAEs for the full $SO(6)$ scalar sector.  In Sec.
\ref{sec:compare} we perform the comparison with the all-loop BAEs.
In Sec.  \ref{sec:discussion} we briefly discuss these results, and
list some interesting related problems which remain unsolved.  In
Appendix \ref{sec:numerical} we present some numerical results which
demonstrate the completeness of our Bethe ansatz solution in the
$SU(3)$ sector for spin chains of short length.

\section{The $SU(3)$ sector}\label{sec:SU3}

We consider in this section, as in Section 4.3.1 of \cite{HM},  the
subsector $SU(3) \subset SO(6)$.

\subsection{Hamiltonian}

In the $SU(3)$ sector, the Hilbert space
is \footnote{Following \cite{HM}, we define the origin of the spin chain at site 0 
instead of site 1.}
\be
\stackrel{\stackrel{0}{\downarrow}}{C^{2}}  \otimes 
\stackrel{\stackrel{1}{\downarrow}}{C^{3}}  \otimes \cdots
\stackrel{\stackrel{L}{\downarrow}}{C^{3}}  \otimes 
\stackrel{\stackrel{L+1}{\downarrow}}{C^{2}} \,.
\label{Hilbertspace}
\ee
The one-loop mixing matrix for operators of the type (\ref{Y0excitation}) is
the open-chain Hamiltonian given by (cf. Eq. (2.15) in \cite{BV})
\be
H = 2g^{2} \left(  Q_{0}^{Y} h_{0, 1} Q_{0}^{Y} +\sum_{l=1}^{L-1} h_{l, l+1} + 
Q_{L+1}^{Y} h_{L, L+1} Q_{L+1}^{Y}  \right) \,,
\label{Hamiltonian}
\ee
where 
\be
g^{2} = \frac{\lambda}{16 \pi^{2}} \,,
\ee
and $\lambda = g^{2}_{YM} N$ is the 't Hooft coupling.
The two-site Hamiltonian $h_{l, l+1}$ is given by 
\be
h_{l, l+1} = \id_{l, l+1} - {\cal P}_{l, l+1} \,,
\label{twosite}
\ee
where $\id$ and ${\cal P}$ are the identity and permutation matrices, 
respectively. The latter can be expressed as
\be
{\cal P} = \sum_{a, b = 1}^{n} e_{a b} \otimes e_{b a} \,,
\label{permutation}
\ee
where $e_{a b}$ is the usual elementary $n \times n$ matrix whose 
$(a, b)$ matrix element is 1, and all others are zero; and here $n=3$.
Moreover, we take $Q^{Y}$ to be the projector \footnote{We choose the 
basis 
\be
|W \rangle = \left( \begin{array}{c}
1\\
0\\
0 \end{array} \right)\,, \quad
|Z \rangle = \left( \begin{array}{c}
0\\
1\\
0 \end{array} \right)\,, \quad
|Y \rangle = \left( \begin{array}{c}
0\\
0\\
1 \end{array} \right)\,. \non 
\ee}
\be
Q^{Y} = \left(  \begin{array}{ccc}
1 & 0 & 0 \\
0 & 1 & 0 \\
0 & 0 & 0 \end{array}  \right) \,,
\label{Qy}
\ee
which is used to implement the restriction (noted below (\ref{Y0excitation}) )
that $Y$ cannot appear at sites 0 and $L+1$. We drop the null rows and columns
of the left and right boundary terms in the Hamiltonian,
\be
H^{L}_{bt} = Q_{0}^{Y} h_{0, 1} Q_{0}^{Y}  \,, \qquad 
H^{R}_{bt} = Q_{L+1}^{Y} h_{L, L+1} Q_{L+1}^{Y} \,.
\label{bts}
\ee
Hence, these boundary terms should be understood as $6 \times 6$
matrices acting on $C^{2} \otimes C^{3}$ and $C^{3} \otimes C^{2}$,
respectively.
We observe that the boundary terms (\ref{bts}) have the symmetry
\be
\left[  H^{L}_{bt} \,,  \h_{0} \g _{1}    \right] = 0 \,, \qquad
\left[  H^{R}_{bt} \,,  \g _{L} \h_{L+1}  \right] = 0 \,,
\label{symmetry1}
\ee
where 
\be
\g = \left( \begin{array}{cc}
                   \h & 0 \\
		    0 & e^{i \theta} \\
	 \end{array} \right)  \,, \qquad 
	 \h \in SU(2)  \,.
\label{g}
\ee	

\subsection{Transfer matrix}

We now proceed to construct the commuting open-chain transfer matrix
which contains the Hamiltonian (\ref{Hamiltonian}).  Following the
general recipe of Sklyanin \cite{Sk}, there are two main ingredients:
$R$-matrices and $K$-matrices.  The $R$-matrix is a solution $R(u)$ of
the Yang-Baxter equation (YBE)
\be
R_{12}(u_{1}-u_{2})\, R_{13}(u_{1})\, R_{23}(u_{2}) =
R_{23}(u_{2})\, R_{13}(u_{1})\, R_{12}(u_{1}-u_{2}) \,.
\label{YBE}
\ee 
For the case at hand with $SU(3)$ symmetry, the $R$-matrix is
well known to be the $9 \times 9$ matrix acting on $C^{3} 
\otimes C^{3}$ given by
\be
R(u) = u \id + i {\cal P} \,.
\label{Rmatrix}
\ee

Here the right $K$-matrix $K^{R}(u)$ is a $6 \times 6$ matrix acting on $C^{3} 
\otimes C^{2}$ which is a solution of the right boundary Yang-Baxter 
equation  (BYBE) \cite{Ch, GZ}
\be
\lefteqn{R_{12}(u_{1}- u_{2})\, K^{R}_{13}(u_{1})\, R_{12}(u_{1}+u_{2})\, 
K^{R}_{23}(u_{2}) }\non \\
& & = K^{R}_{23}(u_{2})\, R_{12}(u_{1}+u_{2})\, K^{R}_{13}(u_{1})\,R_{12}(u_{1}- 
u_{2}) \,.
\label{RBYBE}
\ee 
Assuming that $K^{R}(u)$ has the same symmetry as the right boundary 
term $H^{R}_{bt}$ (\ref{symmetry1}), i.e., 
\be 
\left[  K^{R}(u)  \,, \g  \otimes \h   \right] = 0 \,,
\ee
leads to the ansatz
\be
K^{R}(u) =  \left(  \begin{array}{cccccc}
a_{1}(u)+a_{2}(u) \\
& a_{1}(u) & a_{2}(u) \\
& a_{2}(u) & a_{1}(u) \\
&   &  & a_{1}(u)+a_{2}(u) \\
&   &  &   & a_{3}(u) \\
&   &  &   & & a_{3}(u) \end{array} \right) \,,
\label{KRansatz}
\ee
where matrix elements which are zero are left empty. The boundary 
Yang-Baxter equation (\ref{RBYBE}) together with the regularity 
condition $K^{R}(0) = \id$ imply \footnote{There is in fact a 
one-parameter family of such solutions. We fix the parameter so as to 
match with the boundary term (\ref{Kbtrltn}).}
\be
a_{1}(u) = 1 - u^{2} \,, \qquad a_{2}(u) = - 2i u \,, \qquad a_{3}(u) = 1 + u^{2} 
\,.
\label{KRsltn}
\ee
This $K$-matrix has the feature that its first derivative evaluated at
$u=0$ is proportional to the right boundary term,
\be
\frac{d}{du}K^{R}(u) \Big\vert_{u=0} = 2i \left( H^{R}_{bt}- \id 
\right) \,,
\label{Kbtrltn}
\ee 
up to an additive term proportional to the identity.

The left $K$-matrix $K^{L}(u)$ is a $6 \times 6$ matrix acting on
$C^{2} \otimes C^{3}$ which is a solution of the left BYBE
\be
\lefteqn{R_{12}(-u_{1}+ u_{2})\, K^{L}_{3 1}(u_{1})^{t_{1}}\, 
R_{12}(-u_{1}-u_{2}-\eta)\, 
K^{L}_{3 2}(u_{2})^{t_{2}} }\non \\
& & = K^{L}_{3 2}(u_{2})^{t_{2}}\, R_{12}(-u_{1}-u_{2}-\eta)\, 
K^{L}_{3 1}(u_{1})^{t_{1}}\,R_{12}(-u_{1} +u_{2}) \,,
\label{LBYBE}
\ee 
where $t_{i}$ denotes transposition in the $i^{th}$ space, and 
$\eta = 3 i$ appears in the crossing-unitarity relation
\be
R_{12}(u)^{t_{1}} \, R_{12}(-u-\eta)^{t_{1}} \propto \id \,,
\ee 
where the proportionality factor is some scalar function of $u$.
Assuming that $K^{L}(u)$ has the same symmetry as the left boundary 
term $H^{L}_{bt}$ (\ref{symmetry1}), i.e., 
\be 
\left[  K^{L}(u)  \,, \h  \otimes \g   \right] = 0 \,,
\ee
leads to the ansatz
\be
K^{L}(u) =  \left(  \begin{array}{cccccc}
b_{1}(u)+b_{2}(u) \\
& b_{1}(u) &  & b_{2}(u) \\
&   &  b_{3}(u) \\
&   b_{2}(u) &  & b_{1}(u) \\
&   &  &   & b_{1}(u)+b_{2}(u) \\
&   &  &   & & b_{3}(u) \end{array} \right) \,.
\label{KLansatz}
\ee
We find the following solution of the left BYBE (\ref{LBYBE})
\be
b_{1}(u) = i u + u^{2} \,, \qquad b_{2}(u) = -3 + 2i u \,, \qquad 
b_{3}(u) = -2i u - u^{2} 
\,.
\label{KLsltn}
\ee
This $K$-matrix has the feature that its value at $u=0$ is
proportional to the left boundary term,
\be
K^{L}(0)  = 3 \left( H^{L}_{bt}- \id \right) \,,
\ee 
up to an additive term proportional to the identity.

The transfer matrix $t(u)$ is given by
\be
t(u) = \tr_{a} K^{L}_{0 a}(u)\, T_{a 1 \cdots L}(u) \, K^{R}_{a L+1}(u)\,
\hat T_{a 1 \cdots L}(u)  \,,
\label{transfer}
\ee
where the trace $( \tr )$ is over a 3-dimensional auxiliary space denoted by 
$a$. The argument of the trace acts on 
\be
\stackrel{\stackrel{0}{\downarrow}}{C^{2}}  \otimes 
\stackrel{\stackrel{a}{\downarrow}}{C^{3}}  \otimes
\stackrel{\stackrel{1}{\downarrow}}{C^{3}}  \otimes \cdots
\stackrel{\stackrel{L}{\downarrow}}{C^{3}}  \otimes 
\stackrel{\stackrel{L+1}{\downarrow}}{C^{2}} \,,
\ee
and therefore $t(u)$ acts on (\ref{Hilbertspace}), as does the
Hamiltonian.  The monodromy matrices $T$ and $\hat T$ are given by
\be
T_{a 1 \cdots L}(u) = R_{a 1}(u) \cdots R_{a L}(u) \,, \qquad
\hat T_{a 1 \cdots L}(u) = R_{a L}(u) \cdots R_{a 1}(u) \,.
\label{monodromy}
\ee 
Indeed, it can be shown along the lines \cite{Sk} that the transfer 
matrix (\ref{transfer}) obeys the fundamental commutativity property
\be
\left[ t(u) \,, t(v) \right] = 0 \,,
\label{commutativity}
\ee
by virtue of the fact that the $R$ and $K$ matrices obey their respective YBEs 
(\ref{YBE}), (\ref{RBYBE}), (\ref{LBYBE}). In fact, the latter 
equation for $K^{L}(u)$ was engineered to ensure this commutativity.
It can also be shown 
that this transfer matrix contains the Hamiltonian 
(\ref{Hamiltonian}),
\be
H = c_{1} \frac{d}{du}t(u) \Big\vert_{u=0} + c_{2} \id \,,
\label{tHrelation}
\ee
where
\be
c_{1} = 2 g^{2} \left( \frac{i}{6} (-1)^{L}\right) \,, \qquad 
c_{2} = 2 g^{2}\left(L + \frac{4}{3} \right)\,.
\label{tHcoeffs}
\ee
The relations (\ref{commutativity}) - (\ref{tHcoeffs}) provide 
a direct proof of the integrability of the Hamiltonian.

We observe that the transfer matrix has the $SU(2) \times U(1)$ 
symmetry
\be
\left[ t(u) \,, \h \otimes \g^{\otimes L} \otimes \h \right] = 0 \,,
\ee
where $\g$ and $\h$ are defined in (\ref{g}).
The eigenstates of the transfer matrix therefore form representations 
of $SU(2)$.

\subsection{Bethe ansatz}\label{subsec:BA}

The commutativity property (\ref{commutativity}) implies that it is 
possible to find eigenstates $| \Lambda \rangle$ of the transfer 
matrix $t(u)$ which are independent of $u$,
\be
t(u)\, | \Lambda \rangle = \Lambda(u)\, | \Lambda \rangle \,.
\ee
We now proceed to determine the eigenvalues $\Lambda(u)$ by the 
analytical Bethe ansatz \cite{Re}-\cite{selene}.

Acting with the transfer matrix on the vacuum state (\ref{Y0vacuum}), i.e.,
\be
| Z \cdots Z \rangle = 
\left( \begin{array}{c}
0 \\
1 \end{array} \right) \otimes 
\left( \begin{array}{c}
0 \\
1\\
0 \end{array} \right) \otimes \cdots
\left( \begin{array}{c}
0 \\
1\\
0 \end{array} \right)  \otimes
\left( \begin{array}{c}
0 \\
1 \end{array} \right)  \,,
\ee
we find that the vacuum eigenvalue is given by
\be
\Lambda_{0}(u) = -\frac{(u+i)}{(2u+i)} 
\left[ (2u+3i) (u+i)^{2L+3} + 4(u+i) u^{2L+3} \right]
\,.
\ee
Known results for the closed and open $SU(3)$ chains (see, e.g., 
\cite{KR, dVGR}) suggest that a general eigenvalue should have the 
``dressed'' form
\be
\Lambda(u) &=& -\frac{(u+i)}{(2u+i)} \Bigg\{
(2u+3i)(u+i)^{2L+3} \frac{Q_{1}(u-i/2)}{Q_{1}(u+i/2)} \non\\
&+& u^{2L+3} \left[ B_{1}(u) \frac{Q_{1}(u+3i/2)}{Q_{1}(u+i/2)} 
\frac{Q_{2}(u)}{Q_{2}(u+i)} + B_{2}(u) \frac{Q_{2}(u+2i)}{Q_{2}(u+i)} 
\right] \Bigg\} \,,
\label{eigenval1}
\ee
where
\be
Q_{j}(u) = \prod_{k=1}^{m_{j}} (u - u_{j, k}) (u + u_{j, k}) \,, 
\qquad j = 1\,, 2 \,,
\label{Qs}
\ee
and 
\be
B_{1}(u) + B_{2}(u) = 4(u+i) \,.
\ee
By considering the $L=1$ case, we readily determine
\be
B_{1}(u) = 2u + 3i \,, \qquad
B_{2}(u) = 2u + i \,.
\label{eigenval2}
\ee

The BAEs for the zeros $u_{j, k}$ of the functions $Q_{j}(u)$
(\ref{Qs}) follow the fact that $\Lambda(u)$ in (\ref{eigenval1}) is
analytic at $u=u_{1, k} - i/2$ and at $u=u_{2, k} - i$. In terms of 
the standard notation
\be
e_{n}(u) = \frac{u + i n/2}{u - i n/2} \,,
\label{notation}
\ee
the BAEs take the form
\be
e_{1}(u_{1, k})^{2L+2}  &=& 
\prod_{j=1 \atop j\ne k}^{m_{1}} e_{2}(u_{1, k} - u_{1, j})\, 
e_{2}(u_{1, k} + u_{1, j}) \non \\
& & \times \prod_{j=1}^{m_{2}} e_{-1}(u_{1, k} - 
u_{2, j})\, e_{-1}(u_{1, k} + u_{2, j}) \,, \quad k = 1\,, \ldots 
\,, m_{1} \,, \non \\
1 &=& 
\prod_{j=1 \atop j\ne k}^{m_{2}} e_{2}(u_{2, k} - u_{2, j})\, 
e_{2}(u_{2, k} + u_{2, j})\non \\
& & \times  \prod_{j=1}^{m_{1}} e_{-1}(u_{2, k} - 
u_{1, j})\, e_{-1}(u_{2, k} + u_{1, j}) \,,  \quad k = 1\,, \ldots 
\,, m_{2} \,.
\label{BAEs}
\ee 
As a check, we observe that reducing to the $SU(2)$ subsector by
removing the Bethe roots $\{u_{2, k} \}$ yields exactly the same
result derived in Appendix \ref{sec:SU2}, namely, Eq.  (\ref{BAEs2}).

In view of the relation (\ref{tHrelation}) between the transfer matrix
and the Hamiltonian, we find that the eigenvalues of the Hamiltonian
(\ref{Hamiltonian}) are given by
\be
E = c_{1} \frac{d}{du} \Lambda(u) \Big\vert_{u=0} + c_{2} 
  = 2 g^{2} \sum_{k=1}^{m_{1}} \frac{1}{u_{1, k}^{2} + 1/4} \,.
\label{energy}
\ee 
The problem of determining the one-loop anomalous dimensions of
operators of the type (\ref{Y0excitation}) in the $SU(3)$ sector is in
principle solved by (\ref{BAEs}), (\ref{energy}).

We have verified the completeness of this solution for $L = 1, 2, 3$,
as discussed in Appendix \ref{sec:numerical}.  We observe that for $L
> 1$, the numbers of Bethe roots for a given eigenvalue satisfy
\be
0 \le m_{1} \le L\,, \qquad 0 \le m_{2} \le m_{1} \,.
\ee
Degenerate states $| \Lambda \rangle $ with the same eigenvalue
$\Lambda(u)$ (characterized by given sets of Bethe roots $\{ u_{1, k} 
\}\,, \{ u_{2, k} \}$) do not necessarily form irreducible representations
of $SU(2)$; they are characterized by one or more values of the
$SU(2)$ spin $s$. The following inequality appears to hold
\be
s \le \frac{1}{2}\left(L+2-m_{1}-m_{2}\right) \,.
\ee

\section{The $SO(6)$ sector}\label{sec:SO6}

We have not formulated the transfer matrix for the full $SO(6)$ 
scalar sector, for which the  Hamiltonian is given by Eq. (2.15) in 
\cite{BV}. However, based on our results for the $SU(3)$ and $SU(2)$ 
sectors presented in Sec. \ref{sec:SU3} and Appendix \ref{sec:SU2}, 
respectively, it is not difficult to conjecture the result for the 
BAEs. Indeed, since $SO(6) \approx SU(4)$ has rank three, we expect
that the (one-loop) BAEs are given by 
\be
e_{1}(u_{1, k})^{2L+2} &=& 
\prod_{j=1 \atop j\ne k}^{m_{1}} e_{2}(u_{1, k} - u_{1, j})\, 
e_{2}(u_{1, k} + u_{1, j}) \non \\
& & \times \prod_{j=1}^{m_{2}} e_{-1}(u_{1, k} - 
u_{2, j})\, e_{-1}(u_{1, k} + u_{2, j}) \non \\
& & \times \prod_{j=1}^{m_{3}} e_{-1}(u_{1, k} - 
u_{3, j})\, e_{-1}(u_{1, k} + u_{3, j}) \,, \quad k = 1\,, \ldots 
\,, m_{1} \,, \non 
\ee
\be 
1 &=& 
\prod_{j=1 \atop j\ne k}^{m_{2}} e_{2}(u_{2, k} - u_{2, j})\, 
e_{2}(u_{2, k} + u_{2, j})\non \\
& & \times  \prod_{j=1}^{m_{1}} e_{-1}(u_{2, k} - 
u_{1, j})\, e_{-1}(u_{2, k} + u_{1, j}) \,,  \quad k = 1\,, \ldots 
\,, m_{2} \,, \non 
\ee
\be 
1 &=& 
\prod_{j=1 \atop j\ne k}^{m_{3}} e_{2}(u_{3, k} - u_{3, j})\, 
e_{2}(u_{3, k} + u_{3, j})\non \\
& & \times  \prod_{j=1}^{m_{1}} e_{-1}(u_{3, k} - 
u_{1, j})\, e_{-1}(u_{3, k} + u_{1, j}) \,,  \quad k = 1\,, \ldots 
\,, m_{3} \,. \label{BAESO6}
\ee 
Reducing to the $SU(3)$ subsector by removing the Bethe roots $\{u_{3,
k} \}$ yields (\ref{BAEs}), and reducing further to the $SU(2)$
subsector by also removing the roots $\{u_{2, k} \}$ yields
(\ref{BAEs2}). The open-chain BAEs (\ref{BAESO6}) are essentially 
``doubled'' with respect to the corresponding closed-chain results of 
Minahan and Zarembo \cite{MZ}, except for the exponent on the LHS of 
the first equation, which is $2L+2$ for the open chain with $L+2$ 
sites, and is $L$ for the closed chain with $L$ sites.

\section{Comparison with all-loop BAEs}\label{sec:compare}

All-loop BAEs for the $Y=0$ brane have recently been proposed in an
interesting recent paper by Galleas \cite{Ga}.  We now wish to compare
those equations with the one-loop BAEs which we have proposed for the
scalar sector.  This will require performing the weak-coupling limit
of the former BAEs, and then reducing to the scalar sector.

\subsection{All-loop BAEs}

We begin by recalling the all-loop BAEs from \cite{Ga}: 
\footnote{In the first version of \cite{Ga}, the third set of BAEs 
(here,  (\ref{finalba2})) contained some sign errors, and the 
expression for $\Phi(\lambda)$ differed from (\ref{Phi}).}
\begin{eqnarray}
\label{finalba}
\left[ \frac{x^{+}_{k}}{x^{-}_{k}}  \right]^{(-2L -2N + m_{1} + m_{2} )} 
\Phi(\lambda_{k}) &=& 
\prod_{\stackrel{j=1}{j \neq k}}^{N} \left[ S_{0}(\lambda_{k}, \lambda_{j}) 
S_{0}(\lambda_{j}, -\lambda_{k}) \frac{(x^{-}_{k} + x^{-}_{j})(x^{-}_{k} - x^{+}_{j})}
{(x^{+}_{k} - x^{-}_{j})(x^{+}_{k} + x^{+}_{j})} \right]^2  \nonumber \\
&\times & \prod_{\alpha=1}^{2} \prod_{l=1}^{m_{\alpha}}  
\frac{ (x^{+}_{k} -  z^{-}_{\alpha,l} ) (x^{+}_{k} +  z^{-}_{\alpha,l} )}
{( x^{-}_{k}  - z^{-}_{\alpha,l} )( x^{-}_{k}  + z^{-}_{\alpha,l} ) }
\end{eqnarray}
\begin{eqnarray}
\label{finalba1}
\prod_{j=1}^{N} \frac{(z^{-}_{\alpha, k} + x^{-}_{j})}{(z^{-}_{\alpha, k} - x^{-}_{j})} 
\frac{(z^{-}_{\alpha , k} - x^{+}_{j})}{(z^{-}_{\alpha , k} + x^{+}_{j})} 
\Theta(z^{\pm}_{\alpha , k}) &=&
\prod_{j=1}^{n_{\alpha}}
\frac{(z^{-}_{\alpha , k} + \frac{1}{z^{-}_{\alpha , k}} - \tilde{\lambda}_{\alpha , j} 
- \frac{i}{2g} )}{(z^{-}_{\alpha , k} + \frac{1}{z^{-}_{\alpha , k}} 
- \tilde{\lambda}_{\alpha , j} + \frac{i}{2g} )} 
\frac{(z^{-}_{\alpha , k} + \frac{1}{z^{-}_{\alpha , k}} + \tilde{\lambda}_{\alpha , j} 
- \frac{i}{2g} )}{(z^{-}_{\alpha , k} + \frac{1}{z^{-}_{\alpha , k}} 
+ \tilde{\lambda}_{\alpha , j} + \frac{i}{2g} )} \nonumber \\
&& \;\;\;\;\;\;\;\;\;\;\;\;\;\;\;\;\;\;\;\;\;\;\;\;\;\;\;\;\;\;\;\;\;\;\;\;\; 
\alpha =1,2 \;\;\;\;\;\;\; k=1, \dots , m_{\alpha} \nonumber \\
\end{eqnarray}
\begin{eqnarray}
\label{finalba2}
\prod_{j=1}^{m_{\alpha}} 
\frac{( \tilde{\lambda}_{\alpha , k} -z^{-}_{\alpha ,j} - \frac{1}{z^{-}_{\alpha ,j}} 
+ \frac{i}{2g} )}{( \tilde{\lambda}_{\alpha ,k} -z^{-}_{\alpha ,j} 
- \frac{1}{z^{-}_{\alpha ,j}} - \frac{i}{2g} )} 
\frac{( \tilde{\lambda}_{\alpha ,k} +z^{-}_{\alpha ,j} 
+ \frac{1}{z^{-}_{\alpha ,j}} + \frac{i}{2g} )}{( \tilde{\lambda}_{\alpha ,k} +z^{-}_{\alpha ,j} 
+ \frac{1}{z^{-}_{\alpha ,j}} - \frac{i}{2g} )} &=&
\prod_{\stackrel{j=1}{j \neq k}}^{n_{\alpha}}
\frac{( \tilde{\lambda}_{\alpha ,k} - \tilde{\lambda}_{\alpha ,j}  
+ \frac{i}{g} )}{( \tilde{\lambda}_{\alpha ,k} - \tilde{\lambda}_{\alpha ,j}  - \frac{i}{g} )} 
\frac{( \tilde{\lambda}_{\alpha ,k} + \tilde{\lambda}_{\alpha ,j}  
+ \frac{i}{g} )}{( \tilde{\lambda}_{\alpha ,k} + \tilde{\lambda}_{\alpha ,j}  - \frac{i}{g} )} 
\nonumber \\
&& \;\;\;\;\;\;\;\;\;\;\; \alpha =1,2 \;\;\;\;\;\;\; k=1, \dots , 
n_{\alpha} \,, \nonumber \\ 
\end{eqnarray}
where $\Theta(z^{\pm})$ is given by
\begin{equation}
\label{teta}
\Theta(z^{\pm}) =  \frac{2 z^{+} z^{-}(z^{+} + \frac{1}{z^{+}} 
- \frac{i}{2g} )}{(z^{+} +  z^{-})( z^{+} z^{-} + 1)} \,.
\end{equation}
Moreover, $\Phi(\lambda)$ is given by \cite{Ga}
\begin{equation}
\Phi(\lambda) = \left[\left(\frac{x^{+}}{x^{-}}\right)^{2}
\frac{1}{k^{+}_{0}(-\lambda) k^{-}_{0}(\lambda)}
\right]^2 \,,
\label{Phi}
\end{equation} 
where $S_{0}(\lambda, \lambda')$ and $k_{0}^{\pm}(\lambda)$ are the
scalar factors of the bulk and boundary $S$-matrices, respectively.

Assuming that $z^{\pm}$ satisfy the usual constraint
\be
z^{+} + \frac{1}{z^{+}} - z^{-} - \frac{1}{z^{-}} = \frac{i}{g} \,,
\ee
we find that the quantity (\ref{teta}) simplifies to unity,
\be
\Theta(z^{\pm}) = 1 \,.
\ee
Hence, although the all-loop BAEs seem to depend on both $z^{+}$ and
$z^{-}$, they can in fact be expressed in terms of $z^{-}$ alone.

In order to bring these equations to a more familiar form, we perform 
the following identifications \cite{MM} \footnote{We identify the 
variables $x^{\pm}_{j}$, $z^{-}_{\alpha ,j}$ and $\tilde{\lambda}_{\alpha ,j}$ 
in \cite{Ga} with $x^{\pm}(p_{j})$, $x^{+}(\lambda^{(\alpha)}_{j})$ 
and $\tilde u^{(\alpha)}_{j}$ in \cite{MM}, respectively.}
\be
x^{\pm}_{j} &=& \frac{x^{\pm}_{4, j}}{g} \,, \qquad j = 1, \ldots , 
K_{4} \equiv N \,, \non \\
z^{-}_{1, j}  &=&  \frac{g}{x_{1, j}} \,, \qquad j = 1, \ldots , 
K_{1} \,, \non \\
z^{-}_{1, K_{1}+j}  &=&  \frac{x_{3, j}}{g} \,, \qquad j = 1, \ldots , 
K_{3}\,, \qquad m_{1} \equiv K_{1} + K_{3} \,, \non \\
z^{-}_{2, j}  &=&  \frac{x_{5, j}}{g} \,, \qquad j = 1, \ldots , 
K_{5} \,, \non \\
z^{-}_{2, K_{5}+j}  &=&  \frac{g}{x_{7, j}} \,, \qquad j = 1, \ldots , 
K_{7}\,, \qquad m_{2} \equiv K_{5} + K_{7} \,, \non \\
\tilde{\lambda}_{1 ,j}   &=&  \frac{u_{2, j}}{g} \,, \qquad j = 1, \ldots , 
K_{2} \equiv n_{1} \,, \non \\
\tilde{\lambda}_{2 ,j}   &=&  \frac{u_{6, j}}{g} \,, \qquad j = 1, \ldots , 
K_{6} \equiv n_{2} \,.
\label{identifications}
\ee 
Assuming \footnote{This expression differs from the one given by Eq. (36)
in \cite{MM} by the interchange $j \leftrightarrow k$; however, it
seems to be consistent with the conventions in \cite{Ga}.}
\be
S_{0}(\lambda_{k} \,, \lambda_{j})^{2} = 
\left( \frac{x^{+}_{k}-x^{-}_{j}}{x^{-}_{k}-x^{+}_{j}} \right)
\left( \frac{1-\frac{1}{x^{+}_{j} x^{-}_{k}}}
{1-\frac{1}{x^{-}_{j} x^{+}_{k}}} \right)
\sigma(\lambda_{j}, \lambda_{k})^{2}
\ee
and recalling \cite{HM} 
\be
x^{\pm}(-\lambda) = - x^{\mp}(\lambda) \,,
\ee
the first equation (\ref{finalba}) becomes
\be
\lefteqn{e^{-2i \lambda_{k} (L + K_{4} + 
\frac{K_{1}-K_{3}+K_{7}-K_{5}}{2})} \Phi(\lambda_{k})
= \prod_{\stackrel{j=1}{j \neq k}}^{K_{4}}
\Bigg\{ 
\left( \frac{x^{-}_{4, k} - x^{+}_{4, j}}
{x^{+}_{4, k} - x^{-}_{4, j}}\right)
\left(\frac{1- \frac{g^{2}}{x^{+}_{4, j} x^{-}_{4, k}}}{1- 
\frac{g^{2}}{x^{-}_{4, j} x^{+}_{4, k}}} \right)
\sigma(\lambda_{j}, \lambda_{k})^{2}} \non \\
& & \times 
\left( \frac{x^{-}_{4, k} + x^{-}_{4, j}}
{x^{+}_{4, k} + x^{+}_{4, j}}\right)
\left(\frac{1+ \frac{g^{2}}{x^{-}_{4, j} x^{-}_{4, k}}}
{1+ \frac{g^{2}}{x^{+}_{4, j} x^{+}_{4, k}}} \right)
\sigma(-\lambda_{k},\lambda_{j})^{2} \Bigg\} \non \\
& & \times \prod_{j=1}^{K_{3}}
\left( \frac{x^{+}_{4, k} - x_{3, j}}
{x^{-}_{4, k} - x_{3, j}}\right)
\left( \frac{x^{+}_{4, k} + x_{3, j}}
{x^{-}_{4, k} + x_{3, j}}\right)
\prod_{j=1}^{K_{5}}
\left( \frac{x^{+}_{4, k} - x_{5, j}}
{x^{-}_{4, k} - x_{5, j}}\right)
\left( \frac{x^{+}_{4, k} + x_{5, j}}
{x^{-}_{4, k} + x_{5, j}}\right) \non \\
& & \times \prod_{j=1}^{K_{1}}
\left(\frac{1- \frac{g^{2}}{x_{1, j} x^{+}_{4, k}}}
{1- \frac{g^{2}}{x_{1, j} x^{-}_{4, k}}} \right)
\left(\frac{1+ \frac{g^{2}}{x_{1, j} x^{+}_{4, k}}}
{1+ \frac{g^{2}}{x_{1, j} x^{-}_{4, k}}} \right)
\prod_{j=1}^{K_{7}}
\left(\frac{1- \frac{g^{2}}{x_{7, j} x^{+}_{4, k}}}
{1- \frac{g^{2}}{x_{7, j} x^{-}_{4, k}}} \right)
\left(\frac{1+ \frac{g^{2}}{x_{7, j} x^{+}_{4, k}}}
{1+ \frac{g^{2}}{x_{7, j} x^{-}_{4, k}}} \right) \,, \non \\
& & \qquad\qquad\qquad\qquad k = 1, \ldots , K_{4}\,.
\label{set1}
\ee
With the help of the definitions \cite{MM}
\be
u_{i, j} = x_{i, j} + \frac{g^{2}}{x_{i, j}} \,, \qquad 
i = 1, 3, 5, 7\,,
\label{defn}
\ee
the second set of equations (\ref{finalba1}) becomes
\be
\prod_{j=1}^{K_{4}} \left( 
\frac{1-\frac{g^{2}}{x_{1, k} x^{+}_{4, j}}}
{1-\frac{g^{2}}{x_{1, k} x^{-}_{4, j}}} \right)
\left( 
\frac{1+\frac{g^{2}}{x_{1, k} x^{-}_{4, j}}}
{1+\frac{g^{2}}{x_{1, k} x^{+}_{4, j}}} \right)
= \prod_{j=1}^{K_{2}} \left( 
\frac{u_{1, k}-u_{2,j} - \frac{i}{2}}
{u_{1, k}-u_{2,j} + \frac{i}{2}}\right)
\left( 
\frac{u_{1, k}+u_{2,j} - \frac{i}{2}}
{u_{1, k}+u_{2,j} + \frac{i}{2}}\right) \,, \ k = 1, \ldots , 
K_{1}\,, \non 
\ee 
\be
\prod_{j=1}^{K_{4}} \left( 
\frac{x_{3, k}-x^{+}_{4, j}}{x_{3, k}-x^{-}_{4, j}}\right)
 \left( 
\frac{x_{3, k}+x^{-}_{4, j}}{x_{3, k}+x^{+}_{4, j}}\right)
= \prod_{j=1}^{K_{2}} \left( 
\frac{u_{3, k}-u_{2,j} - \frac{i}{2}}
{u_{3, k}-u_{2,j} + \frac{i}{2}}\right)
\left( 
\frac{u_{3, k}+u_{2,j} - \frac{i}{2}}
{u_{3, k}+u_{2,j} + \frac{i}{2}}\right) \,, \ k = 1, \ldots , 
K_{3}\,, \non 
\ee 
\be
\prod_{j=1}^{K_{4}} \left( 
\frac{x_{5, k}-x^{+}_{4, j}}{x_{5, k}-x^{-}_{4, j}}\right)
 \left( 
\frac{x_{5, k}+x^{-}_{4, j}}{x_{5, k}+x^{+}_{4, j}}\right)
= \prod_{j=1}^{K_{6}} \left( 
\frac{u_{5, k}-u_{6,j} - \frac{i}{2}}
{u_{5, k}-u_{6,j} + \frac{i}{2}}\right)
\left( 
\frac{u_{3, k}+u_{6,j} - \frac{i}{2}}
{u_{5, k}+u_{6,j} + \frac{i}{2}}\right) \,, \ k = 1, \ldots , 
K_{5}\,, \non 
\ee 
\be
\prod_{j=1}^{K_{4}} \left( 
\frac{1-\frac{g^{2}}{x_{7, k} x^{+}_{4, j}}}
{1-\frac{g^{2}}{x_{7, k} x^{-}_{4, j}}} \right)
\left( 
\frac{1+\frac{g^{2}}{x_{7, k} x^{-}_{4, j}}}
{1+\frac{g^{2}}{x_{7, k} x^{+}_{4, j}}} \right)
= \prod_{j=1}^{K_{6}} \left( 
\frac{u_{7, k}-u_{6,j} - \frac{i}{2}}
{u_{7, k}-u_{6,j} + \frac{i}{2}}\right)
\left( 
\frac{u_{7, k}+u_{6,j} - \frac{i}{2}}
{u_{7, k}+u_{6,j} + \frac{i}{2}}\right) \,, \ k = 1, \ldots , 
K_{7}\,. \non \\
\label{set2}
\ee 
Finally, the third set of equations (\ref{finalba2}) becomes 
\be
\lefteqn{
\prod_{j=1}^{K_{1}} \left( 
\frac{u_{2, k}-u_{1,j} + \frac{i}{2}}
{u_{2, k}-u_{1,j} - \frac{i}{2}}\right)
\left( 
\frac{u_{2, k}+u_{1,j} + \frac{i}{2}}
{u_{2, k}+u_{1,j} - \frac{i}{2}}\right)
\prod_{j=1}^{K_{3}} \left( 
\frac{u_{2, k}-u_{3,j} + \frac{i}{2}}
{u_{2, k}-u_{3,j} - \frac{i}{2}}\right)
\left( 
\frac{u_{2, k}+u_{3,j} + \frac{i}{2}}
{u_{2, k}+u_{3,j} - \frac{i}{2}}\right)} \non \\
& & = \prod_{\stackrel{j=1}{j \neq k}}^{K_{2}}
\left( \frac{u_{2, k}-u_{2,j} +i}
{u_{2, k}-u_{2,j} - i}\right)
\left( \frac{u_{2, k}+u_{2,j} +i}
{u_{2, k}+u_{2,j} - i}\right) \,, \ k = 1, \ldots , 
K_{2}\,, \non
\ee
\be
\lefteqn{
\prod_{j=1}^{K_{5}} \left( 
\frac{u_{6, k}-u_{5,j} + \frac{i}{2}}
{u_{6, k}-u_{5,j} - \frac{i}{2}}\right)
\left( 
\frac{u_{6, k}+u_{5,j} + \frac{i}{2}}
{u_{6, k}+u_{5,j} - \frac{i}{2}}\right)
\prod_{j=1}^{K_{7}} \left( 
\frac{u_{6, k}-u_{7,j} + \frac{i}{2}}
{u_{6, k}-u_{7,j} - \frac{i}{2}}\right)
\left( 
\frac{u_{6, k}+u_{7,j} + \frac{i}{2}}
{u_{6, k}+u_{7,j} - \frac{i}{2}}\right)} \non \\
& & = \prod_{\stackrel{j=1}{j \neq k}}^{K_{6}}
\left( \frac{u_{6, k}-u_{6,j} +i}
{u_{6, k}-u_{6,j} - i}\right)
\left( \frac{u_{6, k}+u_{6,j} +i}
{u_{6, k}+u_{6,j} - i}\right) \,, \ k = 1, \ldots , 
K_{6}\,. \label{set3}
\ee
As expected, the open-chain BAEs (\ref{set2}), (\ref{set3}) are 
precisely ``doubled'' with respect to the corresponding
closed-chain results given by Eqs.  (52), (53) in \cite{MM}, 
respectively.
As we shall see below, Eqs.  (\ref{set3}) are not relevant
for the scalar sector.

\subsection{Weak-coupling limit}

We perform the weak-coupling ($g \rightarrow 0$) limit by setting
\be
x_{i, j} \rightarrow \frac{u_{i, j}}{g} \,, \quad i = 1, 3, 5, 7 \,,
\qquad 
x^{\pm}_{4, j} \rightarrow \frac{1}{g} \left(u_{4, j} \pm 
\frac{i}{2}\right) \,,
\ee
and keeping the $u$'s finite.  Since
\be
e^{i \lambda_{k}} = \frac{x^{+}_{4,k}}{x^{-}_{4,k}} \rightarrow 
\frac{u_{4, k} + \frac{i}{2}}{u_{4, k} - \frac{i}{2}} \,,
\ee
the first equation (\ref{set1}) becomes
\be
\left(\frac{u_{4, k} + \frac{i}{2}}{u_{4, k} - \frac{i}{2}} 
\right)^{2L + 2K_{4} + K_{1}-K_{3}+K_{7}-K_{5}} 
\Phi(\lambda_{k})^{-1}
&=&  \prod_{\stackrel{j=1}{j \neq k}}^{K_{4}}
\left( \frac{u_{4, k}-u_{4,j} + i}
{u_{4, k}-u_{4,j} - i}\right)
\left( \frac{u_{4, k}+u_{4,j} + i}
{u_{4, k}+u_{4,j} - i}\right) \non \\
& \times & \prod_{j=1}^{K_{3}}
\left( \frac{u_{4, k}-u_{3,j} - \frac{i}{2}}
{u_{4, k}-u_{3,j} + \frac{i}{2}}\right)
\left( \frac{u_{4, k}+u_{3,j} - \frac{i}{2}}
{u_{4, k}+u_{3,j} + \frac{i}{2}}\right) \non \\
& \times &  \prod_{j=1}^{K_{5}}
\left( \frac{u_{4, k}-u_{5,j} - \frac{i}{2}}
{u_{4, k}-u_{5,j} + \frac{i}{2}}\right)
\left( \frac{u_{4, k}+u_{5,j} - \frac{i}{2}}
{u_{4, k}+u_{5,j} + \frac{i}{2}}\right) \,. \non \\
\ee
Moreover, according to (\ref{Phi}), the factor $\Phi(\lambda_{k})$ 
becomes
\be
\Phi(\lambda_{k})  \rightarrow  
\left(\frac{u_{4, k} + \frac{i}{2}}{u_{4, k} - \frac{i}{2}} \right)^{4} 
\,,
\label{problem}
\ee
assuming $k_{0}^{\pm}(\lambda) \rightarrow  1$, as suggested by 
\cite{BV, HM}.

In the weak-coupling limit, the second set of equations (\ref{set2}) becomes
\be
\prod_{j=1}^{K_{4}} \left( 
\frac{u_{3, k}-u_{4,j} - \frac{i}{2}}
{u_{3, k}-u_{4,j} + \frac{i}{2}}\right)
\left( 
\frac{u_{3, k}+u_{4,j} - \frac{i}{2}}
{u_{3, k}+u_{4,j} + \frac{i}{2}}\right)
= \prod_{j=1}^{K_{2}} \left( 
\frac{u_{3, k}-u_{2,j} - \frac{i}{2}}
{u_{3, k}-u_{2,j} + \frac{i}{2}}\right)
\left( 
\frac{u_{3, k}+u_{2,j} - \frac{i}{2}}
{u_{3, k}+u_{2,j} + \frac{i}{2}}\right) \,, \non
\ee
\be
\prod_{j=1}^{K_{2}} \left( 
\frac{u_{1, k}-u_{2,j} - \frac{i}{2}}
{u_{1, k}-u_{2,j} + \frac{i}{2}}\right)
\left( 
\frac{u_{1, k}+u_{2,j} - \frac{i}{2}}
{u_{1, k}+u_{2,j} + \frac{i}{2}}\right)
= 1 \,, \non
\ee
\be
\prod_{j=1}^{K_{4}} \left( 
\frac{u_{5, k}-u_{4,j} - \frac{i}{2}}
{u_{5, k}-u_{4,j} + \frac{i}{2}}\right)
\left( 
\frac{u_{5, k}+u_{4,j} - \frac{i}{2}}
{u_{5, k}+u_{4,j} + \frac{i}{2}}\right)
= \prod_{j=1}^{K_{6}} \left( 
\frac{u_{5, k}-u_{6,j} - \frac{i}{2}}
{u_{5, k}-u_{6,j} + \frac{i}{2}}\right)
\left( 
\frac{u_{5, k}+u_{6,j} - \frac{i}{2}}
{u_{5, k}+u_{6,j} + \frac{i}{2}}\right) \,, \non
\ee
\be
\prod_{j=1}^{K_{6}} \left( 
\frac{u_{7, k}-u_{6,j} - \frac{i}{2}}
{u_{7, k}-u_{6,j} + \frac{i}{2}}\right)
\left( 
\frac{u_{7, k}+u_{6,j} - \frac{i}{2}}
{u_{7, k}+u_{6,j} + \frac{i}{2}}\right)
= 1 \,.
\ee
The third set of equations (\ref{set3}) remain the same in the 
weak-coupling limit.

\subsection{Reduction to the $SO(6)$ sector}

Following \cite{BS1}, we reduce the above weak-coupling BAEs to the
scalar sector by first transforming to the ``beauty'' form, and then
removing the roots $u_{1, k}\,, u_{2, k}\,, u_{6, k}\,, u_{7, k}$.
The latter procedure corresponds to removing the outer two nodes on
each side of the $su(2,2|4)$ Dynkin diagram, leaving just the three
nodes of $so(6) = su(4)$.  In terms of the notation (\ref{notation}),
it follows that
\be
1 &=&  \prod_{j=1 \atop j\ne k}^{K_{3}} 
e_{2}(u_{3, k} - u_{3, j})\, e_{2}(u_{3, k} + u_{3, j})\non \\
& & \times  \prod_{j=1}^{K_{4}} e_{-1}(u_{3, k} - 
u_{4, j})\, e_{-1}(u_{3, k} + u_{4, j}) \,,  \quad k = 1\,, \ldots 
\,, K_{3} \,, \non 
\ee
\be
e_{1}(u_{4, k})^{2L + 2K_{4} - 4 -K_{3}-K_{5}}  &=& 
\prod_{j=1 \atop j\ne k}^{K_{4}} e_{2}(u_{4, k} - u_{4, j})\, 
e_{2}(u_{4, k} + u_{4, j}) \non \\
& & \times \prod_{j=1}^{K_{3}} e_{-1}(u_{4, k} - 
u_{3, j})\, e_{-1}(u_{4, k} + u_{3, j}) \non \\
& & \times \prod_{j=1}^{K_{5}} e_{-1}(u_{4, k} - 
u_{5, j})\, e_{-1}(u_{4, k} + u_{5, j}) \,, \quad k = 1\,, \ldots 
\,, K_{4} \,, \non 
\ee
\be
1 &=&  \prod_{j=1 \atop j\ne k}^{K_{5}} 
e_{2}(u_{5, k} - u_{5, j})\, e_{2}(u_{5, k} + u_{5, j})\non \\
& & \times  \prod_{j=1}^{K_{4}} e_{-1}(u_{5, k} - 
u_{4, j})\, e_{-1}(u_{5, k} + u_{4, j}) \,,  \quad k = 1\,, \ldots 
\,, K_{5} \,. \label{galleas}
\ee
Comparing these BAEs with those which we have proposed, we see that,
upon identifying $u_{3, k}\,, u_{4, k}\,, u_{5, k}$ in (\ref{galleas})
with $u_{2, k}\,, u_{1, k}\,, u_{3, k}$ in (\ref{BAESO6}),
respectively, the equations almost match.  The only difference is in
the exponent of the term corresponding to the ``massive'' node: in
(\ref{BAESO6}) it is $2L+2$, while in (\ref{galleas}) it is $2L +
2K_{4} - 4 -K_{3}-K_{5}$.  This sort of mismatch also occurs in the
periodic case \cite{MM}.  Evidently, the exact exponent cannot be
deduced from an analysis (such as \cite{Ga}) which is based solely on
$S$-matrices, and requires additional input, such as the weak-coupling
result proposed here.

\section{Discussion}\label{sec:discussion}

We have investigated the open spin chain describing the scalar sector
of the $Y=0$ giant graviton brane at weak coupling.  We have provided
a direct proof of integrability in the $SU(3)$ and $SU(2)$ sectors by
constructing the transfer matrices, namely, (\ref{transfer}) and
(\ref{transfer2}), respectively.  Expanding the transfer matrix $t(u)$
in powers of $u$ generates the Hamiltonian and the higher local
conserved quantities.  We have determined the eigenvalues of these
transfer matrices in terms of roots of the corresponding BAEs, namely,
(\ref{BAEs}) and (\ref{BAEs2}), respectively.  Based on these results,
we have proposed BAEs for the full $SO(6)$ sector (\ref{BAESO6}).
Finally, we have found that, in the weak-coupling limit, the
recently-proposed all-loop BAEs \cite{Ga} essentially agree with
those which we have proposed. 

There are evidently several outstanding questions which remain to be
addressed. It would be interesting to
construct the transfer matrix for the full $SO(6)$ sector and check
directly the proposed BAEs (\ref{BAESO6}).  In that case, the BYBEs
for the $K$-matrices are significantly more complicated than in the
$SU(3)$ case, and we have not yet succeeded to find appropriate
solutions.  Perhaps it may be feasible to investigate other sectors,
as well as finite-size corrections.
Finally, it would also be interesting to investigate the
case of the $Z=0$ brane, for which there are boundary degrees of
freedom.

\section*{Acknowledgments}
I am grateful to C. Ahn, Z. Bajnok and W. Galleas for helpful 
discussions and correspondence.
This work was supported in part by the National Science
Foundation under Grants PHY-0244261 and PHY-0554821.

\appendix

\section{The $SU(2)$ sector}\label{sec:SU2}

We briefly consider here the $SU(2)$ subsector of the $Y=0$ brane. 
This sector consists of only fields $Z$ and $Y$.  The Hamiltonian is again
given by (\ref{Hamiltonian}) - (\ref{twosite}), where the permutation
matrix is given by (\ref{permutation}) with $n=2$.
Choosing the basis
$|Z \rangle = \left({1 \atop 0}\right)$, $|Y \rangle = \left({0 \atop 1}\right)$,
the projector $Q^{Y}$ is now given by  (cf. (\ref{Qy}))
\be
Q^{Y} = \left(  \begin{array}{ccc}
1 & 0 \\
0 & 0 \end{array}  \right) \,.
\ee
The left boundary term can be factorized as follows
\be
H^{L}_{bt} = Q_{0}^{Y} \left(\id - {\cal P}_{0, 1} \right) Q_{0}^{Y} = 
Q_{0}^{Y} (\id -Q_{1}^{Y}) = Q_{0}^{Y} q^{Y}_{1} \,,
\ee
where $q^{Y}$ is defined as \cite{HM}
\be
q^{Y} = \id - Q^{Y} = \left(  \begin{array}{ccc}
0 & 0 \\
0 & 1 \end{array}  \right) \,.
\ee
The right boundary term can be expressed in a similar fashion,
\be
H^{R}_{bt} = q^{Y}_{L} Q_{L+1}^{Y}  \,.
\ee
As noted in \cite{HM},
since the fields at sites $0$ and $L+1$ cannot be $Y$'s, they must be
$Z$'s. That is, the spins at sites $0$ and $L+1$ are fixed, and can 
henceforth be ignored. The Hilbert space is therefore simply (cf. 
(\ref{Hilbertspace}))
\be
\stackrel{\stackrel{1}{\downarrow}}{C^{2}}  \otimes \cdots
\otimes \stackrel{\stackrel{L}{\downarrow}}{C^{2}}  \,,
\ee
and the Hamiltonian becomes
\be
H = 2g^{2} \left(  q^{Y}_{1} +\sum_{l=1}^{L-1} h_{l, l+1} + 
q^{Y}_{L} \right) \,.
\label{Hamiltonian2}
\ee

The Hamiltonian (\ref{Hamiltonian2}) is that of an open XXX
(isotropic) spin-1/2 chain with diagonal boundary terms, which has
been well studied.  The commuting transfer matrix is given by
\cite{Sk}
\be
t(u) = \tr_{a} K^{L}_{a}(u)\, T_{a 1 \cdots L}(u) \, K^{R}_{a}(u)\,
\hat T_{a 1 \cdots L}(u)  \,,
\label{transfer2}
\ee
where the monodromy matrices are as before (\ref{Rmatrix}),
(\ref{monodromy}), and the $K$-matrices are given by
\be
K^{R}(u) =  \left(  \begin{array}{cc}
i + u & 0 \\
0 & i - u \end{array} \right) \,, \qquad
K^{L}(u) =  \left(  \begin{array}{cc}
-2i - u & 0 \\
0 & u \end{array} \right) \,.
\ee
The relation of this transfer matrix to the Hamiltonian 
(\ref{Hamiltonian2}) is again given by (\ref{tHrelation})
where now 
\be
c_{1} = 2 g^{2} \left( \frac{i}{4} (-1)^{L+1}\right) \,, \qquad 
c_{2} = 2 g^{2}\left(L + \frac{1}{2} \right)\,.
\ee

The eigenvalues of the transfer matrix (\ref{transfer2}) are given by
\be
\Lambda(u) = -\frac{2}{(2u+i)} \Bigg\{
(u+i)^{2L+3} \frac{Q_{1}(u-i/2)}{Q_{1}(u+i/2)} 
+ u^{2L+3} \frac{Q_{1}(u+3i/2)}{Q_{1}(u+i/2)}  \Bigg\} \,,
\label{eigenvals}
\ee
where $Q_{1}(u)$ is given by (\ref{Qs}) with $j=1$. Analyticity of 
$\Lambda(u)$ at $u=u_{1, k} - i/2$ leads to the BAEs
\be
e_{1}(u_{1, k})^{2L+2} = 
\prod_{j=1 \atop j\ne k}^{m_{1}} e_{2}(u_{1, k} - u_{1, j})\, 
e_{2}(u_{1, k} + u_{1, j})  \,, \quad k = 1\,, \ldots 
\,, m_{1} \,.
\label{BAEs2}
\ee 
Finally, we note that the eigenvalues of the Hamiltonian
(\ref{Hamiltonian2}) are given in terms of the Bethe roots by
the same relation (\ref{energy}).

\section{Numerical results for the $SU(3)$ sector}\label{sec:numerical}

We have verified the completeness of our Bethe ansatz solution in the
$SU(3)$ sector (\ref{eigenval1}) - (\ref{energy}) for $L = 1, 2, 3$.
Our results for the energies ($E$), spins ($s$), and Bethe 
roots ($\{ u_{1, k}\}$, $\{ u_{2, k}\}$)
are presented in Tables 1 and 2. 

Solving BAEs directly is notoriously difficult.  We have obtained the
Bethe roots by instead using ``McCoy's method'' (see, e.g., \cite{ADM,
FM, NR}).  The basic idea is to work backwards: one first explicitly
computes the eigenvalues $\Lambda(u)$ as polynomials in $u$ by
numerically diagonalizing the transfer matrix; and then one solves the
$T-Q$ equations (\ref{eigenval1}) for $Q_{j}(u)$, which are also
polynomials in $u$.  Finally, one finds the zeros of $Q_{j}(u)$, which
are the sought-after Bethe roots $u_{j, k}$.  This method therefore
produces solutions of the BAEs without actually solving the equations!
Although this method is inconvenient for determining the eigenvalues of
an integrable Hamiltonian (direct diagonalization is much faster), it
is ideal for determining all the Bethe roots -- and therefore checking
the completeness of a Bethe ansatz solution -- for small values of $L$.

We have verified that the energies, computed from the Bethe roots
using (\ref{energy}), coincide with the result obtained by direct
diagonalization of the Hamiltonian (\ref{Hamiltonian}).

For each value of $SU(2)$ spin $s$, there is a $(2s+1)$-fold degeneracy.
Taking into account this degeneracy, we find in total $2^{2} 3^{L}$
levels for each value of $L$, which coincides with the dimension of
the Hilbert space (\ref{Hilbertspace}).  In this way, we verify the
completeness of the Bethe ansatz solution.

Finally, it may be worth noting that the tables reveal many
``accidental'' degeneracies: levels described by different sets of
Bethe roots having the same energy.  This suggests that the
Hamiltonian may have some interesting higher symmetry.

\begin{table}[h!] 
  \centering
  \begin{tabular}{|c|c|c|c|c|}\hline
    $L$ & $E/(2g^{2})$ & $s$ & $\{ u_{1, k} \}$ & $\{ u_{2, k} \}$\\
    \hline
     1  &  0  &  3/2  & -- & --\\
        &  1  &  1/2  & $\sqrt{3}/2$ & 0 \\
	&  2  &  1  & 1/2  & --\\
	&  2  &  0  & $ \pm i/2$ & 0 \\
	&  3  &  1/2  & $1/(2\sqrt{3})$ & 0 \\
     2  &  0  &  2  & -- & -- \\
        &  0.585786 & 1 & 1.20711 & 0 \\
	&  1        & 3/2 & $\sqrt{3}/2$ & -- \\
	&  1.26795  & 0, 1/2 & $0.716015 \pm 0.512521  i$  & 0 \\
	&  2  &  1  & $1/2$ & 0 \\
	&  2  &  1  & $ \pm i/2$ & -- \\
	&  2  &  0, 1/2  & $ \pm i/2$ & 0 \\
	&  3  &  3/2  & $1/(2\sqrt{3})$ & -- \\
	& 3.41421 & 1 & 0.207107 & 0 \\
	&  4  &  1/2  & $1/(2\sqrt{3})\,, \sqrt{3}/2$ & $\sqrt{2/3}$ \\
	& 4.73205 & 0, 1/2 & 0.230955, 0.668326 & 0 \\
    \hline
   \end{tabular}
   \caption{Energy, spin, and Bethe 
   roots for $L = 1\,, 2$.}
   \label{table:L12}
\end{table}
\newpage
\begin{table}[h!] 
  \centering
  \begin{tabular}{|c|c|c|c|c|}\hline
     $E/(2g^{2})$ & $s$ & $\{ u_{1, k} \}$ & $\{ u_{2, k} \}$\\
    \hline
          0  &  5/2  & -- & --\\
         0.381966 & 3/2 & 1.53884 & 0 \\
	 0.585786 & 2 & 1.20711 & -- \\
	 0.82259  & 1/2, 1 & $1.11504 \pm 0.545054 i$ & 0 \\
	 1.07919  & 0 & $0.709462\,, 0.695742 \pm 0.987839 i$ & 
	$0\,, 1.56857 i$ \\
	 1.26795  & 3/2 & $0.716015 \pm 0.512521 i$ & -- \\
	 1.38197  & 3/2 & 0.688191 & 0 \\
	 1.38197  & 1/2 & $\pm  i/2\,, 1.36676 i$ & 0 \\
	 1.58579  & 1/2, 1 & $0.513712 \pm 0.499602 i$ & 0 \\
	 1.69722  & 1/2 & $\pm  i/2\,, 1.88488 i$ & 0 \\
	 2        & 2 & 1/2 & -- \\
	 2        & 1, 3/2 & $\pm i/2$ & -- \\
	 2        & 0, 1 & $\pm i/2$ & 0 \\
	 2.58579  & 1 & $0.5\,, 1.20711 $ & 1.0505 \\
	 2.61803  & 3/2 & 0.363271 & 0 \\
	 3        & 0 & $0.661848\,, 0.531587 \pm 0.501604 i$ & 
	$0\,, i/\sqrt{2}$ \\
	 3        & 1/2 & $\pm i/2 \,, \sqrt{3}/2$ & 1 \\
	 3.31526  & 0 & $0.665753\,, 0.26039 \pm 0.500009 i$ & $0\,, 
	1.15861 i$ \\
	 3.32164  & 1/2, 1 & $0.396968\,, 0.949669$ & 0 \\
	 3.41421  & 2 & 0.207107 & -- \\
	 3.61803  & 1/2 & $\pm i/2 \,, 0.606658$  & 0 \\
	 3.61803  & 3/2 & 0.16246  & 0 \\
	 4        & 1 & $(\pm 1 +\sqrt{2})/2$ & 1 \\
	 4.41421  & 1/2, 1 & $0.17238\,, 0.970407 $ & 0 \\
	 4.68474  & 0 & $0.221275\,, 0.667076 \pm 0.508802 i$ & 
	$0\,, 0.810943 i$ \\
	 4.73205  & 3/2 & $0.230955\,, 0.668326$ & -- \\
	 5        & 1/2 & $\pm i/2 \,, 1/(2\sqrt{3})$ & $\sqrt{5}/3$ \\
	 5        & 0 & $0.212263\,, 0.478697 \pm  0.501676 i$ & 
	$0\,, i \sqrt{3/2}$ \\
	 5.30278  & 1/2 & $\pm i/2 \,, 0.229729$ & 0 \\
	 5.41421  & 1 & $0.207107\,, 0.5$ & 0.62964 \\
	 5.85577  & 1/2, 1 & $0.179337\,, 0.427295$ &  0 \\
	 6.92081  & 0 & $0.176138\,, 0.409803\,, 0.883877$ & $0\,, 
	0.678531$ \\
         \hline
   \end{tabular}
   \caption{Energy, spin, and Bethe 
   roots for $L = 3$.}
   \label{table:L3}
\end{table}
%


\strut

\newpage
\strut

\end{document}